\documentclass[10pt,conference,letterpaper,twoside]{IEEEtran}
\usepackage{times}
\usepackage{url}

\usepackage{amssymb}
\usepackage{amssymb}
\usepackage{amstext}
\usepackage{amsmath}
\usepackage{setspace}
\usepackage{mdwlist}

\usepackage{subfigure}
\usepackage[pdftex]{graphicx}
\DeclareGraphicsExtensions{.pdf,.gif}

\usepackage{setspace}

\begin{document}
\title{Maximizing Cloud Providers Revenues via\\Energy Aware Allocation
Policies}

% \author{\IEEEauthorblockN{Michael Shell}
% \IEEEauthorblockA{School of Electrical and\\Computer Engineering\\
% Georgia Institute of Technology\\
% Atlanta, Georgia 30332--0250\\
% Email: http://www.michaelshell.org/contact.html}
% \and
% \IEEEauthorblockN{Homer Simpson}
% \IEEEauthorblockA{Twentieth Century Fox\\
% Email: homer@thesimpsons.com}
% \and
% \IEEEauthorblockN{James Kirk\\ and Montgomery Scott}
% \IEEEauthorblockA{Starfleet Academy\\
% San Francisco, California 96678-2391\\
% Telephone: (800) 555--1212\\


\author{\IEEEauthorblockN{Michele Mazzucco\IEEEauthorrefmark{1}\IEEEauthorrefmark{3},
Dmytro Dyachuk\IEEEauthorrefmark{2} and
Ralph Deters\IEEEauthorrefmark{2}}
\IEEEauthorblockA{\IEEEauthorrefmark{1}University of Cyprus, Cyprus}
\IEEEauthorblockA{\IEEEauthorrefmark{2}University of Saskatchewan, Canada}
\IEEEauthorblockA{\IEEEauthorrefmark{3}University of Tartu, Estonia}}


\maketitle

%\begin{document}

%\title{Maximizing Cloud Providers Revenues via Energy Aware Allocation
%Policies}

%\thispagestyle{empty}

%\date{\today}
%\begin{abstract}
%\end{abstract}

%\section*{\centering Abstract}
%{\em 
%\noindent blablabla
%}
{
\abstract
%A server farm is examined, where a number of servers are offered for rent to
%paying customers. 
%Dynamic allocation policies
%aiming at satisfying the conflicting goals of maximizing the users' experience
%while minimizing the energy consumption are discussed. 
%In this paper we look at the scenarios where Platform-as-a-Service providers offer
%their computational resources to be leased. 
Cloud providers, like Amazon, offer their data centers' computational and storage capacities for lease to
paying customers. 
High electricity consumption, associated with running a data center, not only reflects on its carbon footprint, but also increases the costs of running the data center itself.
%In this paper we advocate a method of dynamically switching servers on and off. 
%The objective of the proposed method is to
This paper addresses the problem of maximizing the revenues 
of Cloud providers by trimming down their electricity costs. 
As a solution allocation policies which are based on
the dynamic powering servers on and off are introduced and evaluated. The
policies aim at satisfying the conflicting goals of maximizing the users' experience while minimizing the amount of consumed electricity.
%A model for estimating the amount of energy needed to run servers under 
%different workloads is presented, and 
The results of numerical experiments and
simulations are described, showing that the proposed scheme performs well under different
traffic conditions.
\endabstract
}

%\section{TODO LIST}
%D: We will remove this part as soon as we will be done with proofreading ;)

%\listoftodos

\section{Introduction}
\label{sec:introduction}

In recent years large investments have been made to 
build data processing centers, purpose-built facilities composed of
thousands of servers and providing storage and computing services within and 
across organizational boundaries. 
%These massive architectures provide advantages for both users and 
%service providers, as their size makes it possible to achieve substantial 
%economies of scale that are simply not possible for `in-house' server farms:
%for example, the cost for people shifts from top to nearly
%irrelevant~\cite{hamilton:2009}. 
%This enables organizations to lease the computing infrastructure they require, 
%rather than purchase a whole server cluster, as providers 
%such as Google or Amazon sell CPU time or storage space on their servers at 
%very competitive prices through on-line Web Services.
Whether used for scientific or commercial purposes, the energy and ecological 
costs (apart from the electricity, a typical data center drawing 15 MW 
of power consumes about 1,400 cubic meters of water per 
day~\cite{hamilton:2009a}) required to operate these 
computing platforms has already reached very high values, {\it e.g.}, in 2006,
data centers used 1.5\% of all the electricity produced in the 
US~\cite{pedram:2009}. 
Apart from the carbon footprint, the high energy consumption negatively 
affects the cost of computations itself, especially in the presence of the 
constantly growing price for electricity\footnote{http://www.eia.doe.gov/}.

Nowadays, it is becoming clear that the next logical step
in the development of data centers is building `green' data centers, {\it
i.e.},  data centers that are energy efficient. Currently most researchers are 
focusing on optimizing the energy efficiency on the hardware level. 
Also, a lot of similar research has been done in the area of power constrained
mobile and portable computing devices, such as laptops, smartphones, PDAs,
etc. However, another method, which has not been studied to the same extent,
is based on dynamic turning on and off servers `on demand'.
In the context of Cloud providers, which offer services like Platform-as-a-Service (PaaS), it is important 
to ensure its stable operation, which eventually will lead to building a reputation
of a dependable PaaS provider.
%What in its turn will attract more customers and thus increase the profits. PaaS consumers rely on the service in order to expand their computational capacities.  
Thus, for the PaaS providers it is important to meet customers' requirements in
terms of both availability and performance. Unfortunately, there is no easy
solution to this problem, as a large portion of expenses for running a data
center is constituted by electricity costs. Therefore, Cloud providers are facing the problem of choosing the right 
number of servers to run in order to avoid
over-provisioning,  as it is a major contributor to excessive power consumption, while meeting availability and
performance requirements.

%Thus, in this paper we study the latter method in the context Cloud providers which offer their %computational resources to be leased. 

%Achieving high utilization, {\it i.e.}, useful work per dollar invested, is very important to
%cut power consumption and make the service provision more profitable. 
%Unfortunately, the typical CPU utilization of servers in Internet utilities ranges between 15\% and %35\%~\cite{andrzejak:2002, hamilton:2008}, 
%while the utilization of any resource in excess of 50\% occurs very
%rarely~\cite{stewart:2007}.

In this paper we propose and evaluate energy-aware allocation policies that aim to maximize the
average revenue received by the provider per unit time. This is achieved by
improving the utilization of the server farm, {\it i.e.}, by powering excess servers off. 
The policies we propose are based on $(i)$ dynamic estimates of user demand, and
$(ii)$ models of system behaviour. 
The emphasis of the latter is on generality rather than analytical tractability. 
Thus, we use some approximations to handle the resulting models.
However, those approximations lead to algorithms that perform well under
different traffic conditions and can be used in real systems. 

The rest of the paper is organized as follows. Relevant related work is
discussed in Section~\ref{sec:related_work}. Section~\ref{sec:cloud} describes
the system model. The mathematical analysis and the resulting policies for 
server allocation are presented in Section~\ref{sec:policies}. 
Section~\ref{sec:ServerPowerUsageEstimation} introduces a model for estimating
the amount of power consumed by servers under different loading conditions, 
while a number of experiments where the allocation policies are compared under 
different traffic conditions are reported in Section~\ref{sec:experiments}. 
Finally, Section~\ref{sec:conclusions} concludes the paper.

\section{Related Work}
\label{sec:related_work}

In the last decade researchers have started to focus on improving the power
consumption of computer and communication systems. 
%The problem of energy efficiency in mobile devices and laptops
%has already attracted a lot of attention. 
However, the problem of data centers energy efficiency is relatively new. 
All the efforts in this area can be categorized in the following way:
\begin{itemize*}
\item Intensive -- optimizing power consumption of a server, {\it e.g.}, by
means of managing CPU voltage/frequency;
\item Extensive -- minimizing power consumption for a server pool, {\it e.g.},
by switching servers on/off;
\item Hybrid -- combining the intensive and extensive methods together.
\end{itemize*}
 
%$(i)$ intensive,  
%$(ii)$ extensive, and 
%$(iii)$ hybrid, a combination of the previous two methods.

Most of the intensive approaches have tried to 
minimize the power consumption when the number of servers is fixed. 
While Google engineers have called for systems designers to develop servers 
that consume energy in proportion to the amount of computing work they 
perform~\cite{barroso:2007} and Microsoft engineers have been working on 
better power management on the operating system layer~\cite{microsoft:2009}, 
servers still consume as much as 65\% of their peak power when
idle~\cite{greenberg:2009}. 
Elnozahy {\it et al.}~\cite{Elnozahy:2002} and Sharma {\it et
al.}~\cite{Sharma:2003}
investigated the potential benefits of scaling down the CPU voltage/frequency  (and
consequently power consumption) according to the offered
load. The results showed that savings can be as big as 20-29\%.

As for extensive approaches, most of the research considered scenarios where
the number of running servers can be controlled at runtime. Thus, the server
farm's energy requirements are reduced by switching some servers off whenever 
it is justified by demand conditions. 
Changes in the pool size are made in a reactive
and/or proactive manner. Reactive methods change the size of the server pool according 
the changes in the load, while proactive algorithms try to determine
the number of the servers beforehand using demand forecasting 
mechanisms~\cite{chen:2005,Hedwig:2009}.

Running too many servers increases the electricity consumption, as even in
the idle mode the servers consume a significant amount of electricity. 
On the other hand, having too few servers switched on requires running those
servers' CPUs at higher frequencies, which consequently increases the energy usage. 
Therefore, hybrid approaches ({\it e.g.},~\cite{Elnozahy:2002,
chen:2005}) attempt to find a rational tradeoff between the number of servers
switched on and the voltage/frequency of the CPU on each server.

Another approach which stands out, as opposed to the previously discussed
ones, was proposed by Qureshi et al. ~\cite{qureshi:2009}. In that paper, the
authors address the problem of minimizing the electricity costs in Content
Delivery Networks (CDN). Given that CDNs have their content replicated 
in each CDN center and the price for electricity varies depending on the
geographical region and time, the authors propose to dynamically re-routing
incoming traffic to the locations with the lowest electricity prices.

\section{The Model}
\label{sec:cloud}

%Next, we model a business model for cloud providers. 
%Requests for CPU reservation arrive according to an independent Poisson process
%with rate $\lambda$.

The provider has a cluster of $S$ identical processors/cores ({\it
servers}, from now on), $n$ running and $(S - n)$ switched off. 
The provider offers each server for a lease, and a customer who rents a server
({\it e.g.}, by running a virtual machine on it) is essentially creating a job.
The size of the job is the length of the lease, and since the client decides
when to terminate the lease, the job size is not known a priori.
Servers are not shared, so each server can handle at maximum one job at any given time (as it will be described in Section~\ref{sec:ServerPowerUsageEstimation}, since the power drained by each CPU is a linear function of the load, the model we propose here can be applied to a scenario where multiple {\it virtual} machines are running on a physical CPU).
If, once a server has finished processing a request, no other jobs enter the system, the server begins to idle ({\it i.e.}, it consumes energy without generating any revenue).

\begin{figure}[ht!]
\centering
\includegraphics[width=0.42\textwidth]{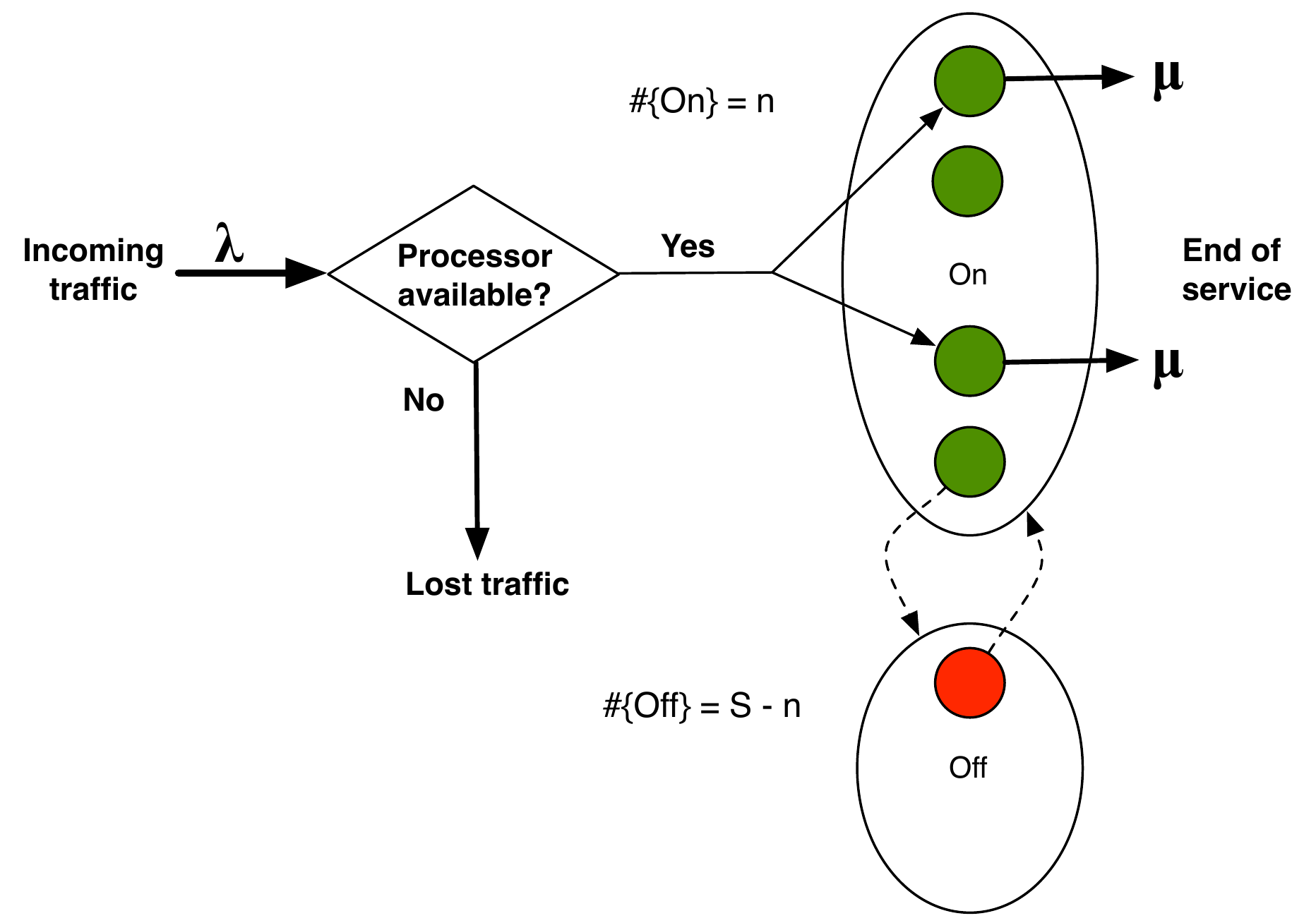}
\caption{System model for cloud providers.}
\label{fig:model_amazon}
\end{figure}

The contract that regulates the provisioning contract states, among the other
things, that for each job \emph{a user pays a charge which is proportional to
the job size}, while the cost the provider bears for running a server is
$c$ \$ per unit time.
%The contract that regulates the provisioning contract includes, among the
%others, the following clause:
%
%\begin{theorem}[Charge]
%\begin{it}
%For each job, a user pays a charge which is proportional to the
%job size.
%\end{it}
%\end{theorem}
Determining the amount of charge is outside the scope of this paper.
Besides, this could also include the charges related to the use of 
storage space or network bandwidth.
%\begin{it}
%The cost for using a server is $c$ \$ per unit time.
%\end{it}
Finally, an arrival finding all $n$ servers busy is blocked and lost,
without affecting future arrivals, see Figure~\ref{fig:model_amazon}, while
running servers consume energy, which costs $r$ \$ per kWh.

%\vspace{-0.3cm}

Within the control of the provider is the `resource allocation' policy,
which decides how many servers to run.
The objective is to find the optimal number of servers, $n$, that should be
switched on in order to optimize the provider's profit. The extreme values,
$n=0$ and $n=S$, correspond to switching respectively off, or on, all available
servers. 

Unfortunately, because of the random nature of user demand, static policies
would under-perform, as servers would be under-utilized when the traffic is
low -- wasting
energy and reducing the provider's revenues -- and overloaded during peak hours,
 missing the profit opportunities.
In order to tackle these issues the provider should be
able to dynamically change the number of running servers in response to changes
in user demand. The problem is how to do that in a sensible manner.

During the intervals between consecutive policy invocations,
the number of running servers remains constant.
Those intervals, which will be referred to as `observation windows', are used 
by the controlling software to collect traffic statistics and obtain 
current estimates of the average arrival rate ($\lambda$) and service
time ($1/ \mu$) as well as the squared coefficients of variation of the above
values (the variance divided by the square of the mean), $ca^{2}$ and $cs^{2}$
respectively.
These values are used by the allocation policy at the next
decision epoch.

It is assumed that the time it takes to change the state of a server is negligible compared to the size of the observation windows. 
That assumption let us neglect the amount of time/energy wasted by servers during reconfigurations. 
Moreover, in a practical implementation, a decision to switch a server off does
not necessarily have to take effect immediately. If a job is being served at
that time, it is allowed to complete before the server is turned off.

{\bf N.B} The assumption that the power up/down operations are instantaneous can be relaxed, at the expenses of complicating the allocation policy. 
We deliberatively opted  not to do so as introducing a short power up/down interval has a little effect on the optimal number of servers to run. On the other hand, if the time it takes to power a server up/down is about the same as the configuration interval ({\it i.e.}, 10 and 30 minutes), than the energy wasted during system reconfigurations should be explicitly taken into account. 

While different metrics can be used to measure the performance of a computing 
system, as far as the service provider is concerned, the performance of the 
system is measured by the average revenue, $R$, earned per unit time. 
That value can be estimated as

\begin{equation}
R = \frac{c}{\mu}T - rP \mbox{,}
\label{eq:amazon_variable}
\end{equation}

\noindent where $c / \mu$ is the average charge paid by a customer for having
his/her job run, $T$ is the system's throughput, and $P$ is the total average
power consumed by the running servers (servers that are currently switched off
do not consume anything).

Please note that, although we make no
assumption regarding the relative magnitudes of charges and costs parameters, the 
most challenging case is when they are close to each other. 
If the charge for
executing a job is much higher than the cost paid by the provider to run a
server, one could guarantee a positive (but not optimal) revenue by switching 
on all servers, regardless of the load. 
On the other hand, if the charge is smaller than 
the cost, than it would be better to switch all
servers off.
Finally, the above model can be easily extended in a number of different ways. For example,
one might include the cost for tearing servers up and down, as well as the cost 
for a smaller mean time between failures (MTBF) of the hardware. However, it is important to 
to note that the proposed approach can be used in scenarios when the price for electricity is not
constant and depends on the time of the day, week, month, etc. In this case, during each
reconfiguration a different value for $c$ should be used.

\section{Policies}
\label{sec:policies}

In order to develop a meaningful framework for energy consumption control, it
is necessary to have a quantitative model of user demand and service provision.
Assuming that jobs enter the system according to an independent Poisson
process with rate $\lambda$, we model the number of jobs inside the system,
for a fixed number of servers $n$, as the number of jobs in an Erlang loss (or
Erlang-B) system with $n$ trunks and traffic intensity $\rho = \lambda / \mu$.

%\begin{equation}
%\rho = \frac{\lambda}{\mu} \mbox{.}
%\label{eq:offered_load}
%\end{equation}

Thus, we can treat the resulting
system as an $M/GI/n/n$ queuing model (the `M' stands for Markovian arrivals),
which has independent and identically distributed (i.i.d.) service times with
a general distribution (the `GI') and independent of the arrival
process, $n$ servers, and no extra waiting spaces ({\it e.g.}, if all servers are busy, further jobs
are lost), augmented with the economic parameters introduced in
Section~\ref{sec:cloud}. Since the Erlang-B model is insensitive  to the
distribution of job sizes, we do not need to worry about the distribution of job lengths. 
In other words, the blocking probability is independent of the service time distribution beyond its mean;
thus, the state probabilities of this system are the same as that of the
corresponding purely Markovian $M/M/n/n$ system where the service times are 
exponentially distributed. 
This model ignores the time-dependence sometimes 
found in job arrival processes. However, this time-dependence often tends to be
not too important over short time intervals. 
%Moreover, even though the  choices
%made may be sub-optimal when the simplifying assumption about the distribution 
%of interarrival intervals does not hold, they should nevertheless lead to 
%reasonable allocation policies.
% there is no queue, and thus any job that finds all the servers busy is lost.
%Under suitable assumptions about the nature of user demand, it is possible to
% evaluate explicitely the effect of particular server allocation on the 
% achievable revenues. That algorithm is sufficiently fast to be performed
% on-line.  Moreover, it can be implemented in any system, but it might lose
% its optimality if the simplyfying assumptions about the nature of user demand
% are not satisfied.
%The simplification consists of assuming that jobs arrive according to an
% independent Poisson process with rate $\lambda$. Service times, instead, are 
% allowed to have general distribution with finite squared coefficient of
% variation. %and their required service times are distributed exponentially
% with mean $1/\mu$.

%In this paper, we focus on highly loaded server farms. 
When $\rho = n$, the system is critically loaded in the limit, and is said to be
in the Quality and Efficiency-Driven (QED) regime, also known as Halfin-Whitt
regime~\cite{halfin:1981}. 
%When
%$\rho > n$, the system is overloaded, and is said to be in the
%Efficiency-Driven (ED) regime. 
In this paper, we focus on heavily loaded server farms where $\rho
\sim n$,  as our aim is to switch off servers in excess while serving as many customers as 
possible. Moreover, we assume that the number of running servers increases 
if the arrival rate grows, {\it i.e.}, $n \rightarrow \infty$ as $\lambda
\rightarrow \infty$, while the service time distribution does not change with
$n$. Under these circumstances, there is a clear separation of 
time scales~\cite{whitt:2002}: as $n$ increases, arrivals and completions occur
more and more quickly ({\it i.e.}, in a fast time scale), while the experience
of individual jobs does not change ({\it i.e.}, in a slow time scale).

Under the Erlang loss model, the number of jobs inside
the system can be modeled as a Birth-and-Death process with a finite state
space, $\{0, 1, \ldots, n\}$. 
An arriving job that finds $j$ ($j < n$) jobs
being served causes a transition to state $(j+1)$ at rate
$\lambda_{j}=\lambda$. 
A completing job at state $j$ ($j=1, \ldots, n$) 
causes a transition to state $(j-1)$ at rate $\mu$, and thus jobs leave the 
system at rate $\mu_{j}= j\mu$. %, see Figure~\ref{fig:markov_chain_amazon}.
%
%
%
%\begin{figure*}[htb!]
%\centering
%\includegraphics[width=0.5\textwidth]{figures/markov_chain_amazon}
%\caption{State transition diagram for the model depicted in
%Figure~\ref{fig:model_amazon}.}
%\label{fig:markov_chain_amazon}
%\end{figure*}
%
%
%The stationary distribution of the number of jobs present may be found by
% solving the balance equation:
%
Denote by $p_{j}$ the stationary probability that there are $j$ jobs in the
$M/GI/n/n$ queue, $j=0,1, \ldots, n$. 
%The balance across the cuts leads to the
%recursion
%
%\begin{equation}
%p_{j}= \frac{\rho}{j} p_{j-1} \mbox{ ,}
%\label{eq:Erlang_B_balance_recursive}
%\end{equation}
%
%\noindent whose general solution can be expressed in the form
After some algebraic manipulations, the balance across the cuts can be expressed in the form 
 
\begin{equation}
p_{j} = \frac{\rho^{j}}{j!} p_{0} \mbox{ .}
\label{eq:Erlang_B_balance_general}
\end{equation}

Steady-state for this Birth-and-Death process 
%sketched in Figure~\ref{fig:markov_chain_amazon} 
exists if, and only if,
Equation~\eqref{eq:Erlang_B_balance_general} can be norma\-lized, {\it i.e.}, if 
%$\displaystyle{\sum^{n}_{j=0} p_{j}= 1}$. 
$\sum^{n}_{j=0} p_{j}= 1$.
Under this model, the steady-state always exists, and from the 
normalization condition, we obtain~\cite{mitrani:1998}
 
\begin{equation}
p_{0} = \left[\sum^{n}_{j=0} \frac{\rho^{j}}{j!} \right]^{-1} \mbox{.}
\label{eq:mmnn_p0}
\end{equation}

The probability of losing a job, {\it i.e.}, the probability $p_n$ to be in
state $n$, is given by the Erlang-B formula

\begin{equation}
%p_{n} = \frac{\frac{\rho^{n}}{n!}}{\sum^{n}_{j=0} \frac{\rho^{j}}{j!}}
%
%p_{n} = \frac{\displaystyle{\frac{\rho^{n}}{n!}}
%}{\displaystyle{
%\sum^{n}_{j=0} \frac{\rho^j}{j!}
%}}
%p_{n} = \frac{\rho^{n}}{n!} \left[\sum^{n}_{j=0} \frac{\rho^{j}}{j!}
% \right]^{-1}
p_{n} = B(n, \rho) = \frac{\rho^{n}}{n!} p_{0} \mbox{ .}
%=  \frac{\rho^{n}}{n!} \left[\sum^{n}_{j=0} \frac{\rho^{j}}{j!} \right]^{-1}
\label{eq:ErlangB}
\end{equation}

Because of the factorial and
large power elements, Equation~\eqref{eq:ErlangB} is very difficult to
calculate directly from its right-hand side when $n$ and $\rho$ are large. However, it can be computed efficiently using the
following iterative scheme~\cite{hudousek:2003}

\begin{equation}
\left \{
\begin{array}{l}
B(0,\rho) = 1\\
B(n+1, \rho) = \displaystyle{\frac{\rho B(n, \rho)}{n+1 +\rho B(n, \rho)}}
\end{array} \right. \mbox{.}
\label{eq:ErlangB_recursive}
\end{equation}

If the arrival process is not Poisson, then the insensitivity property is lost,
and the appropriate queueing model becomes $G/GI/n/n$, for which there is no
exact solution. However, an acceptable approximation for the blocking probability is provided 
by the formula (see Whitt,~\cite{whitt:1984a})

\begin{equation}
p_{n} = B \left(\frac{n}{z}, \frac{\rho}{z} \right) \mbox{,}
\label{eq:erlangB_approx}
\end{equation}

\noindent where $z$ is the asymptotic peakedness of the arrival process, 
defined as the variance divided by the mean of the steady-state queue length
in the asso\-cia\-ted $G/GI/\infty$ model (see~\cite{whitt:1984a} for more 
details). That value can be computed using the following formula

\begin{equation}
z = 1 + (ca^{2} - 1) \eta {\mbox ,}
\label{eq:erlangB_peakedness}
\end{equation}

\noindent where $\eta$ is defined as

\begin{equation}
\eta = \mu \int_{0}^{\infty} [1 - G(t)]^{2} dt {\mbox ,}
\label{eq:erlangB_eta}
\end{equation}

\noindent and $G(t)$ is the
cumulative distribution function (CDF) of the service time distribution with 
mean $1 / \mu$ and variance $\sigma_{s}^{2}$.
%After some manipulation with the Gauss error function, we obtain
%\begin{equation}
%G(t) = \frac{1}{2} [ 1 + erf(\frac{x - 1/ \mu}{\sigma \sqrt{2}}) ]
%\end{equation}

Given the limited amount of information
available, evaluating $G(t)$ is very challenging. Thus, we distinguish between
three cases:

Case 1: $ca^{2} = 1$. The interarrival intervals are exponentially distributed,
and $z$ evaluates to 1. Thus, Equation~\eqref{eq:erlangB_approx} reduces to
Equation~\eqref{eq:ErlangB}.

Case 2: $ca^{2} \neq 1$ and $cs^{2} = 1$. The service times are exponentially
distributed, $\eta = 1/2$ and therefore $z$ is

\begin{equation}
z = 1 + \frac{(ca^{2} - 1)}{2} \mbox{.}
\end{equation}

Case 3: $ca^{2} \neq 1$ and $cs^{2} \neq 1$. We use a normal
approximation %~\cite{jennings:1996} 
to solve
Equation~\eqref{eq:erlangB_peakedness}. Denoted by $N(m, \sigma^{2})$ a normal random variable with mean $m$ and variance $\sigma^{2}$. 
We approximate the distribution $G(t)$ by the distribution of
$N(1/ \mu, \sigma_{s}^{2})$, and compute the integral in
Equation~\eqref{eq:erlangB_eta} using the Legendre-Gauss integration
method.

Finally, since the service time distribution might change over the time, it
may be convenient to periodically recompute the peakedness factor. Denoted by
$z_{k}$ the peakedness at decision epoch $k$. At time $(k+1)$, the new
peakedness can be estimated as

\begin{equation}
z_{k+1} = 1 + (z_{k} - 1) \eta_{k+1} / \eta_{k} \mbox{.}
\end{equation}

Having defined the stationary distribution of the number of jobs present, the
average number of jobs entering the system (and completing service) per unit
time is

\begin{equation}
T = \lambda(1 - p_{n}) \mbox{,}
\label{eq:throughput_Erlang_B}
\end{equation}

\noindent with $(1 - p_{n})$ being the probability that an incoming job finds
an idle server.

The above expressions, together with~\eqref{eq:ErlangB_recursive}, enable the
average revenue $R$ to be computed efficiently and quickly, {\it e.g.},
$B(100000, \rho)$ can be evaluated in about 0.2 seconds using an Intel Core Duo processor.
When that is done for different set of parameter values, it becomes clear that
$R$ is a unimodal function of $n$, {\it i.e.}, it has a single maximum, which
might be $n=S$, or $n=0$, see discussion in Section~\ref{sec:cloud} (this does not depend on the assumption that the electricity cost is constant over the time). 
We do not have a
mathematical proof of this proposition, but have verified it in several numerical experiments. 
Since the cost
for evaluating $R$ is domininated by the computation of
Equations~\eqref{eq:ErlangB_recursive} and~\eqref{eq:erlangB_eta} (where
the latter has to be computed only once), one can search
for the optimal number of servers to run by evaluating $R$ for consecutive values of 
$n$, stopping either when $R$ starts decreasing or, if that does not happen, 
when the revenue increase becomes smaller than some value $\epsilon$. This can be
justified by arguing that the revenue is a concave function with respect to $n$. 
Intuitively, the economic benefits of switching on more servers become less and
less significant as $n$ increases. On the other hand, the loss of potential 
revenues become more and more significant as $n$ decreases. Such behavior is 
an indication of concavity. One can therefore assume that any local maximum 
reached is, or is close to, the global maximum.

%\subsection{Simpler Allocation Heuristics}

The allocation policy described above, which will be refferred to as `Optimal'
policy, requires the evaluation of Equations~\eqref{eq:ErlangB_recursive}
and~\eqref{eq:erlangB_eta}. It may therefore be desirable to have simpler heuristics that allow decisions to
be taken faster and with less information.

\subsection{Adaptive Heuristic} 
\label{sec:adaptive_heuristic}

Deciding on the number of servers to run requires to balance between the server  farm's
utilization and service quality (availability).
High utilization is typically obtained at the
cost of lower availability. 
%frequent and long waiting times. 
Therefore, it is a common belief that
high utilization and good service quality can not coexist.
However, the behaviour of large server farms working in QED
regime differs from that of Kingman's Law ({\it i.e.}, delays/job losses are
very common under heavy load) in that service quality is carefully balanced with server efficiency.

Thus, we propose the following `Adaptive' heurisitc. From the statistics
collected during a window, estimate the arrival rate, $\lambda$, and average 
service time, $1/\mu$. For the duration of the next window, allocate the servers according to

\begin{equation}
	\label{eq:qed_policy}
    n =  \left \lceil \rho + \beta \sqrt{\rho} \right \rceil \mbox{,}
\end{equation}

\noindent where the quantity $\beta \sqrt{\rho}$ is used for dealing with 
stochastic variability, and $-1 \le \beta \le 1$.

\subsection{Predictive Heuristic}
\label{sec:predictive_heuristic}

One can observe that the previously discussed policies simply
adapt to the changes in user demand by assuming that the traffic during the next window will be the
same as that of the previous window. The realism of that assumption can be
disputed, as the load typically follows certain
patterns (daily, weekly, etc.). 
Thus, is might be desirable to design a policy
which tries to predict the user demand.

Denoted by $\rho_{k}$ is the estimated load at window $k$. Instead of simply
adapting to the observed load and assuming that  $\rho_{k} = \rho_{k + 1}$, 
one can try to forecast and estimate what $\rho_{k+1}$ will be using the 
historical data. 
%Since we are dealing with large server farms, it seems reasonable to assume
%that the load is stationary, {\it i.e.}, variance and
%autocorrelation structure do not change over time, or change very slowly.
Thus, a simple
and efficient heuristic using a double exponential smoothing to
estimate the future arrival rate can be employed. For any time period
$k$, the smoothed value $S_{k}$ is found by solving the following
system of equations

\begin{equation}
\left \{
\begin{array}{l}
S_{k} = \alpha \lambda_{} + (1 - \alpha) (S_{k - 1} + b_{k - 1}) \\ 
b_{k} = \gamma(S_{k} - S_{k - 1}) + (1 - \gamma) b_{k - 1}\\
\end{array}
\right. \mbox{.}
\end{equation}

%\noindent with $\alpha$ and $\gamma$ being two constants in the interval $0,
%\ldots, 1$, and $b_{k -1}$ being the trend from the previous period. In the
%above scheme, recent observations are given relatively more weight in 
%forecasting than older ones. 
The first equation adjusts the
smoothed value $S_{k}$ adding $b_{k - 1}$ to the last smoothed value, $S_{k
- 1}$, while the second equation updates the trend.
%\end{enumerate*}
%Different schemes can be employed to initialize the algorithm; here we
%use
%
%\begin{equation}
%\left \{
%\begin{array}{l}
%S_{1} = \lambda_{1} \\
%b_{1} = \displaystyle{\frac{\lambda_{n} - \lambda_{1}}{n-1}}
%\end{array}
%\right. \mbox{.}
%\end{equation}
In this work we used the least squared method in order to find the best values
for $\alpha$ and $\gamma$.
% {\it i.e.}
%, those
%reducing the mean squared error, are computed using non-linear optimization
%techniques.
 %the Levenberg–-Marquardt algorithm.

Having computed the smoothed and the trend values at time $k$, the forecast for
the arrival rate at time $(k+1)$, $\lambda_{k+1}^{F}$, is computed as
%\begin{equation}
$\lambda_{k+1}^{F} = S_{k} + b_{k} \mbox{.} $
%\end{equation}

\section{Server Power Usage Estimation}
\label{sec:ServerPowerUsageEstimation}

The amount of electricity drawn by servers depends on several factors.
Moreover, realistic cost models should take into account wasted
energy such as power conversion losses and the power used for cooling
purposes.
Different algorithms can be employed to estimate the energy requirements of a
data center, the simplest one assuming that the power usage of a server
is constant, while the most complex
models using also disk metrics gathered from some operating system tools such
as  {\it iostat}, in addition to the CPU utilization, or performance 
counters~\cite{rivoire:2008}.

Since most of the Cloud applications are most likely web applications, we 
conducted an experiment aiming at finding the dependency
between the energy consumption and CPU utilization for a common web application. 
In order to avoid biased applications, {\it i.e.}, with high CPU consumption
per job, we have chosen Wordpress\footnote{Wordpress is a popular open source
application which implements a blog, see http://wordpress.org/}  as a study
case. This application runs on top of the {\it LAMP} stack
({\it i.e.}, Linux, Apache, MySQL and PHP),
and thus represents a significant fraction of
the applications running not only in the Cloud but in the Internet as well. 
Moreover, Wordpress jobs are not completely CPU bound, as the application
uses a database as a backend, whose operations are I/O bound. 
Unfortunately, the default configuration of LAMP is far from being optimized for 
 high throughput under heavy load. Therefore, we had to
 perform a number of tune ups, including installing XCache -- which
 caches the compiled PHP code, thus preventing re-compiling the same code for
 every arrival -- and tuning the TCP stack, the Linux kernel and the Apache
 configuration.
 
  %Also, some tuning at the networking level was necessary, with
 %the {\it TIME\_WAIT}, {\it KEEPALIVE} and {\it KEEPALIVE\_PROBES} settings of
 %the TCP stack being lowered. In the Linux settings we increased the number of
 %allowed file descriptors per user to 10,000. To prevent memory thrashing by
 %keeping too many idle processes on a server we lowered HTTP's Keep-Alive to 5
 %seconds, so each process after 5 seconds of inactivity drops current
 %connection and picks the next from the waiting queue. Finally, as the size of
 %the waiting queue was enlarged to 10,000 by changing the backlog setting of
 %TCP.

The server was hosted on a machine with Dual Xeon Dual Core CPUs running at 
2.8 Ghz, 2 Gb of RAM, 7200 RPM hard drive and 1 Gbps network card. 
The power consumption was measured every minute in the presence of an 
increasing workload, which was  generated by Tsung 1.3.2\footnote{http://tsung.erlang-projects.org/}. 
The workload consisted of clients arriving according to a Poisson process at
an increasing rate. Each client replayed a prerecorded session, which included
checking the front page, browsing the posts with some specific tags, as well
as searching in the blog. Surprisingly, most of the HTTP requests were serving
dynamic content, as the static content, which consisted from CSS files and 
JavaScript libraries, was cached on client's side after the first client's request.

\begin{figure}[ht!]
\centering
\includegraphics[width=0.48\textwidth]{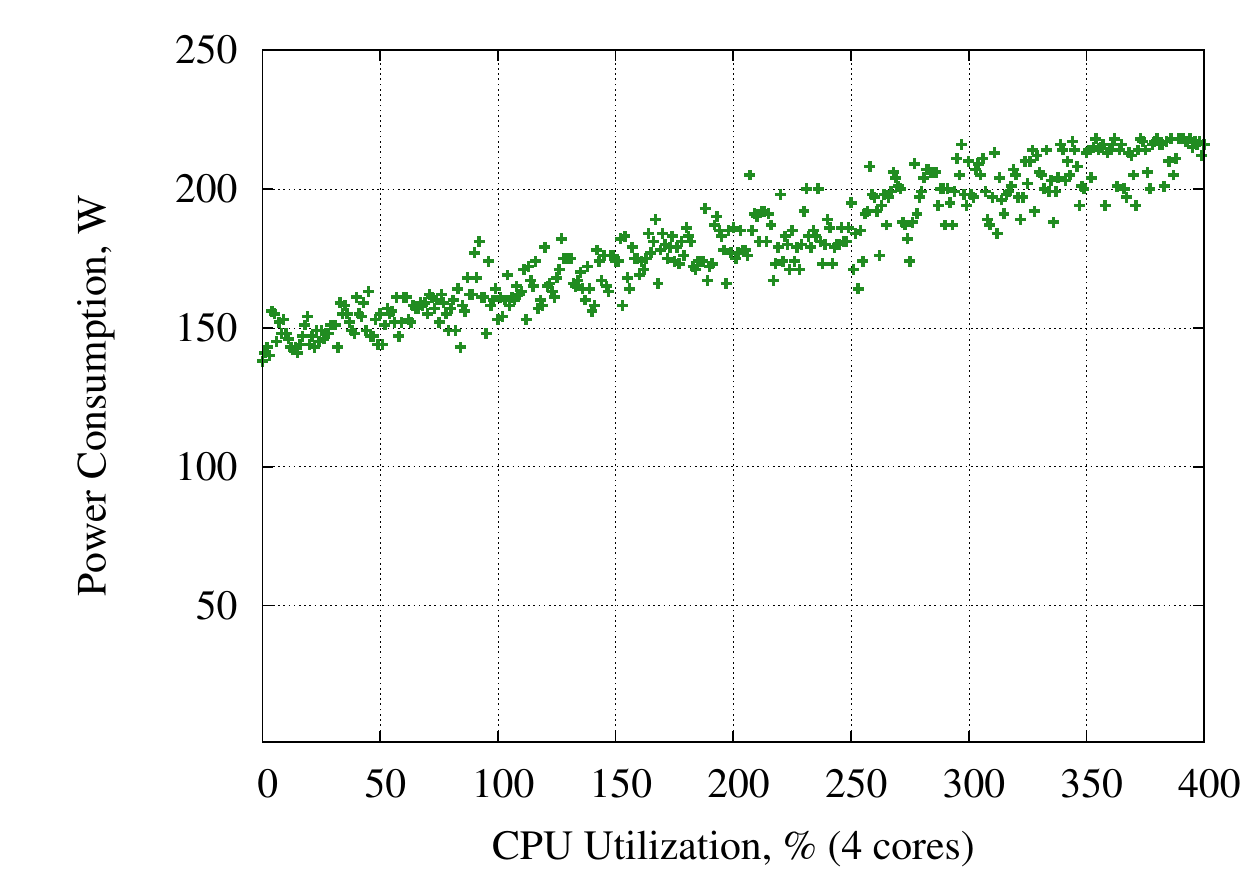}
\caption{Measured energy consumption.}
\label{fig:energy_consumption}
\end{figure}

Figure \ref{fig:energy_consumption} demonstrates the relationship between
power consumption and CPU utilization. In the idle mode, the energy 
consumption stayed at the steady 140 W. 
As shown in Figure \ref{fig:energy_consumption}, the power
consumption grows linearly with the increase of CPU utilization. 
Noise in the power consumption can be attributed 
to noise in the CPU utilization due the irregularity in the request traffic. 
Besides, the fluctuations in the CPU utilization require dynamic
usage of the cooling fans, which in turn amplifies the fluctuations in the
power consumption. 
The power consumption peeks at 220 W when the CPU utilization reaches
values higher than 375\%.
Due to the lack of space, we do not present the behavior of the response time, 
which stayed under one second for loads up to 70\%.

%From the above observations we believe that a common web application
%running on top of LAMP uses up 70\% of the CPU power, while the power
%consumption of server is approximately a linear function of the CPU
%utilization (see also~\cite{fan:2007} and~\cite{barroso:2007}).

Therefore, the average power consumed by a data center per unit time,
$P$, can be estimated as

\begin{equation}
\label{eq:power_linear}
P = n e_{1} + \bar{m}(e_{2} - e_{1}) 
\mbox{,}
\end{equation}

\noindent where $e_{1}$ is the energy consumed per unit time by idle
servers, $e_{2}$ is the energy drawn by each busy server, and $\bar{m}$ is the
average number of servers running jobs ($\bar{m} \leq n$)

\begin{equation}
\bar{m} = \left \lceil \frac{T}{\mu} \right \rceil \mbox{.}
\end{equation}

We have carried out some tests with some other models, and found that the
estimate given by Equation~\eqref{eq:power_linear} gets within 10\% of the
one using performance counters.
% and thus we use the linear
%model for the purposes of this study.

%\input{sec_internet_utilities.tex}

%\input{sec_utilitity_functions.tex}

%\input{sec_allocation_policies.tex}

%\input{sec_numerical_simulations.tex}
\section{Performance Evaluation}
\label{sec:experiments}

Various experiments were carried out, with the aim of evaluating how the
proposed policies affect the maximum achievable
revenues. We assume a server farm with a Power Usage Effectiveness (PUE) of
1.7~\cite{greenberg:2009}. The PUE is one of
the metrics used to measure the efficiency of data centers, and 
it is computed as the ratio between the total facility power and the IT
equipment power. Also, to reduce the number of variables, if not otherwise stated, the
following features and assumptions were held fixed:

\begin{itemize}
\item The data center is composed of 25,000 machines, configured as in
Section~\ref{sec:ServerPowerUsageEstimation}. Therefore, $S = 100,000$.
\item The power consumption of each Xeon machine ranges
	between 140 and 220 W, see Figure~\ref{fig:energy_consumption}. In other 
	words, each server ({\it e.g.}, core, fans, disk and network interface) has a 
	direct consumption between 35 and 55 W.
	Since the server farm has a PUE factor of 1.7, the minimum and maximum power
	consumption are approximately $e_{1}$ = 59 and $e_{2}$ = 94 W per
	server.
  \item The cost for electricity, $r$, is 0.1 \$ per kWh~\cite{neb:2009}.
  \item The average job size, $1 / \mu$, is set to 50 minutes.
\item Completed jobs generate an amount of income of 0.085 \$/hour. Charges are
proportional to the job length, and therefore each job is worth on average
0.071 \$.
\item Jobs are not completely CPU bound. Instead, when a server is busy, the
average CPU utilization is 70\%. In other words, busy servers draw~69.58
Wh, and thus each job costs, for electricity, 0.0058 \$ on average.
\end{itemize}

To make the results more realistic we
take indirect costs  into account as well. These include the cost of capital and
equipment amortization (servers as well as power generators,
transformers, UPS systems, etc.), and account for twice the cost of consumed
electricity.

\begin{figure}[h!]
\centering
\includegraphics[width=0.48\textwidth]{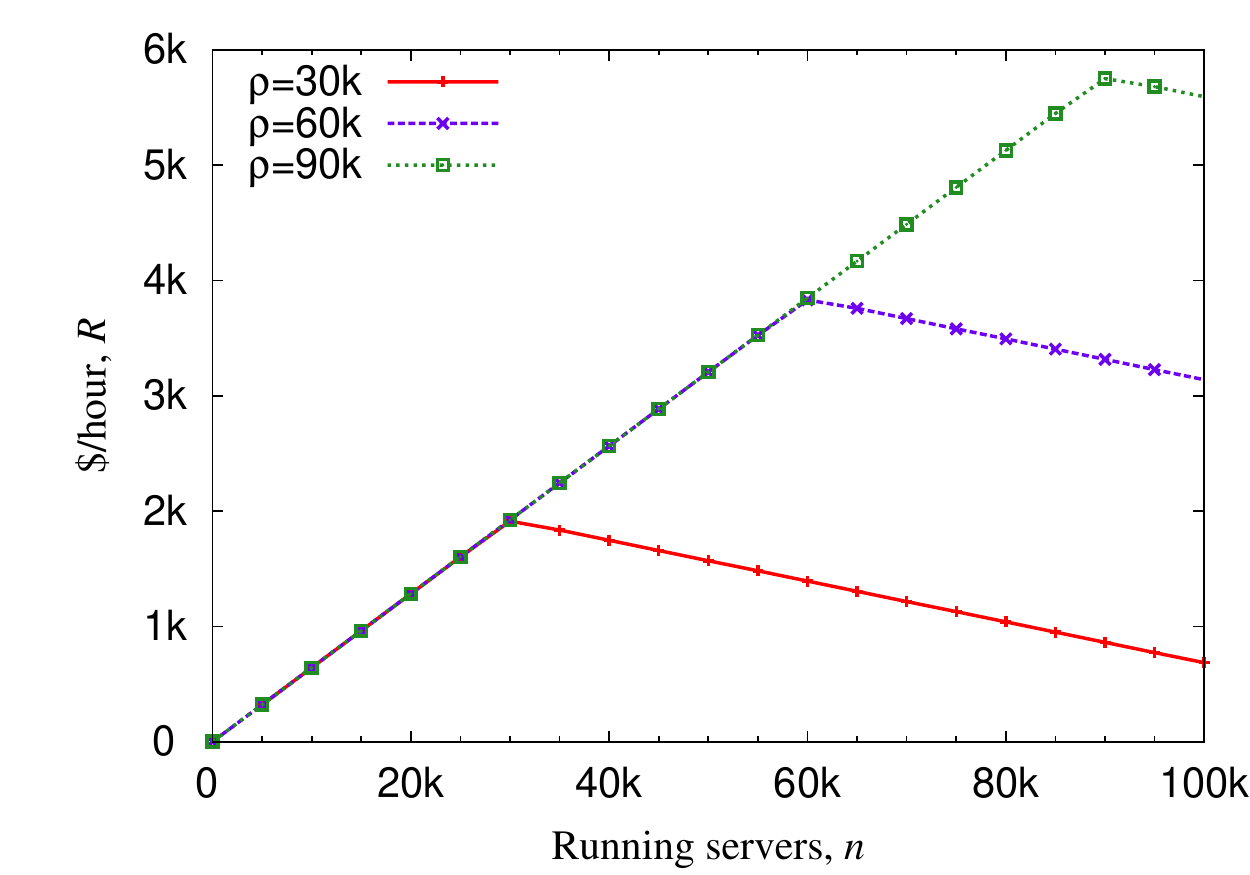}
\caption{Revenue as function of the running servers.}
\label{fig:erlangB_exp1}
\end{figure}

The first experiment is purely numerical.
In Figure~\ref{fig:erlangB_exp1} we examine how the number of running servers
affects the average earned revenue per unit time under different loading
conditions.
The potential offered load is increased from 30\% to 90\% by
increasing the rate at which new jobs enter the system, from 36,000 to 108,000
jobs per hour.

The figure illustrates the following points:
\begin{enumerate}
	\item In each case there is an optimal number of servers that should be
	switched on;
	\item The heavier is the load, the higher is the optimal number of servers as well as the maximum achievable revenue;
	\item When $n > n_{opt}$, the system under-performs because the cost of
  	running idle servers erodes revenues;
  	\item When $n < n_{opt}$, the system under-performs because it misses
  	potential revenues.
\end{enumerate}

Next, we evaluate the performance of the proposed policies via event-driven
simulation.
For comparison reasons, two versions of the `Static' policy, a policy which runs
always the same amount of servers, is also displayed. One runs $n
= S/2 = 50,000$ servers, while the other $n = S = 100,000$. 
We vary the load between 5\% and 99.5\% by varying the arrival rate, {\it i.e.},  $\lambda =
6,000, \ldots, 119,400$ jobs/hour. 
Each point in the figure represents one run lasting 264 hours ({\it i.e.}, 11
days), while reconfigurations occur every 2 hours. During each run, between
1.6 (low load) and 35 million (high load) jobs arrive into the system
(the number of jobs admitted into the system is a bit smaller under heavy load).
Samples of achieved revenues are collected approximately every 24 hours and are
used at the end of each run to compute the corresponding 95\% confidence
interval, which is calculated using the Student's t-distribution.

\begin{figure}[ht!]
\centering
\includegraphics[width=0.48\textwidth]{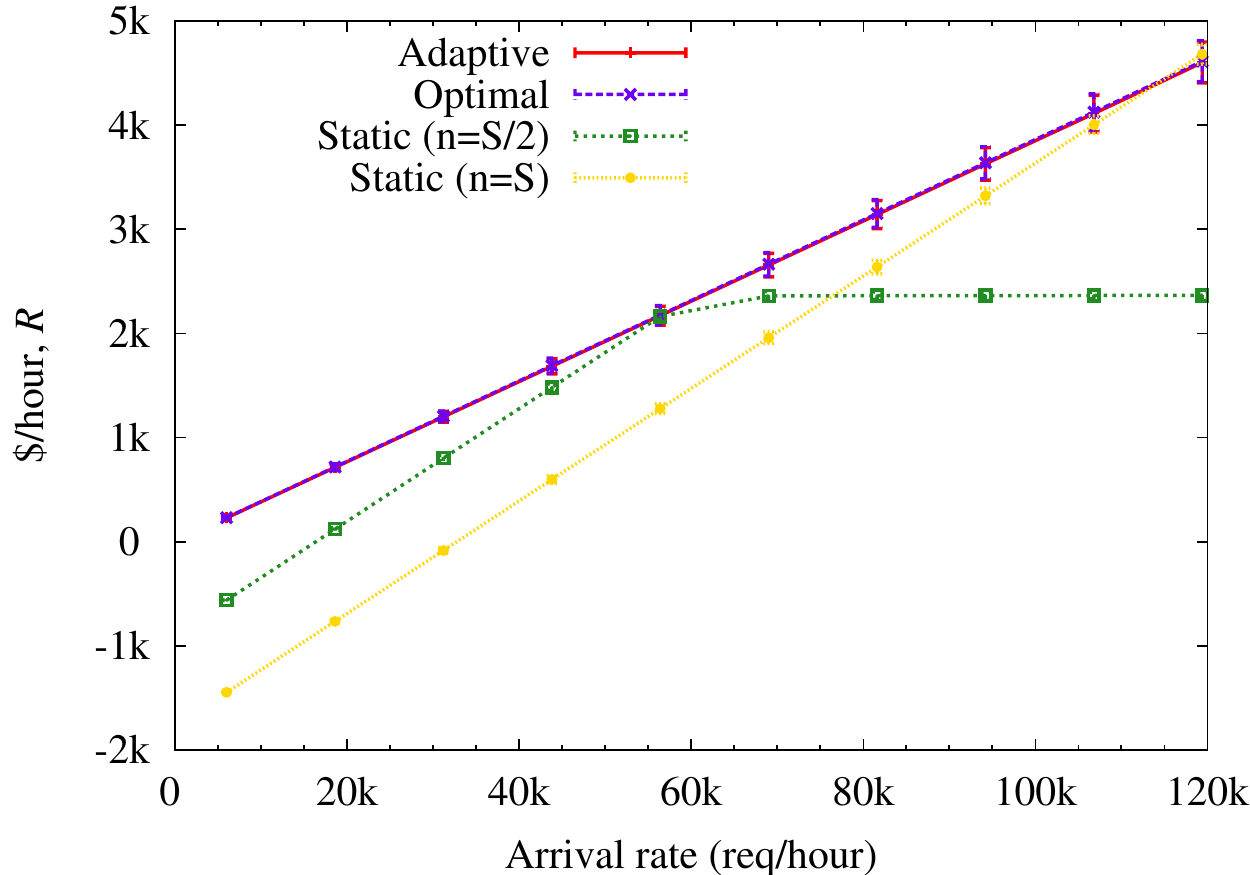}
\caption{Observed revenues for different policies. Markovian scenario.}
\label{fig:exponential}
\end{figure}

The most notable feature of the graph plotted in Figure~\ref{fig:exponential}
is that the performance of the `Static' policies produce negative
revenues under light load (because of the servers running idle), while the one
with parameter $n=S/2$ performs poorly when the load increases,
because too many jobs are lost. On the other hand, the `Adaptive' heuristic
(with parameter $\beta = 0.2$) produces revenues that grow with the offered 
load, and almost as high as those obtained by the more computationally
expensive `Optimal' algorithm. This suggests that the `Adaptive' heuristic
might be a suitable choice for practical implementation.

\begin{figure}[ht!]
\centering
\includegraphics[width=0.48\textwidth]{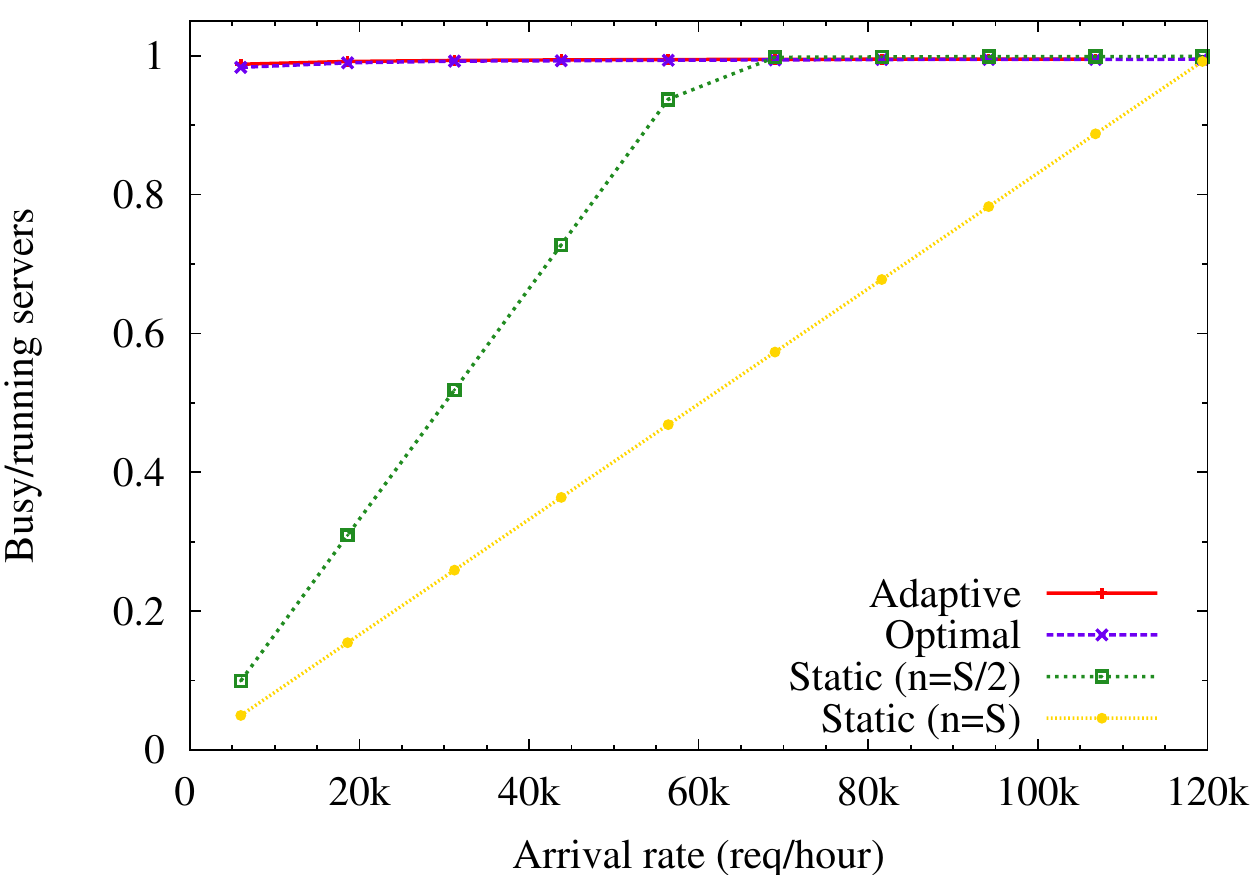}
\caption{Ratio between busy and running servers.}
\label{fig:ratio}
\end{figure}

Figure~\ref{fig:exponential} does not allow to see a comprehensive picture, but it
shows that the policies we propose perform better than the static ones, it
does not provide any insight about the optimality of the algorithms.
Therefore, in Figure~\ref{fig:ratio} we show the ratio between busy and running
servers: a value close to 1 means that the policy performs very well, while a
value close to 0 means that the algorithm does not behave properly. 
The figure shows that the `Adaptive' heuristic is always
very close to 1. 
%Also, from the information we have logged, 
The percentage of 
lost jobs obtained by those policies is always very low, thus ensuring a good
user experience.

\begin{figure}[ht!]
\centering
\includegraphics[width=0.48\textwidth]{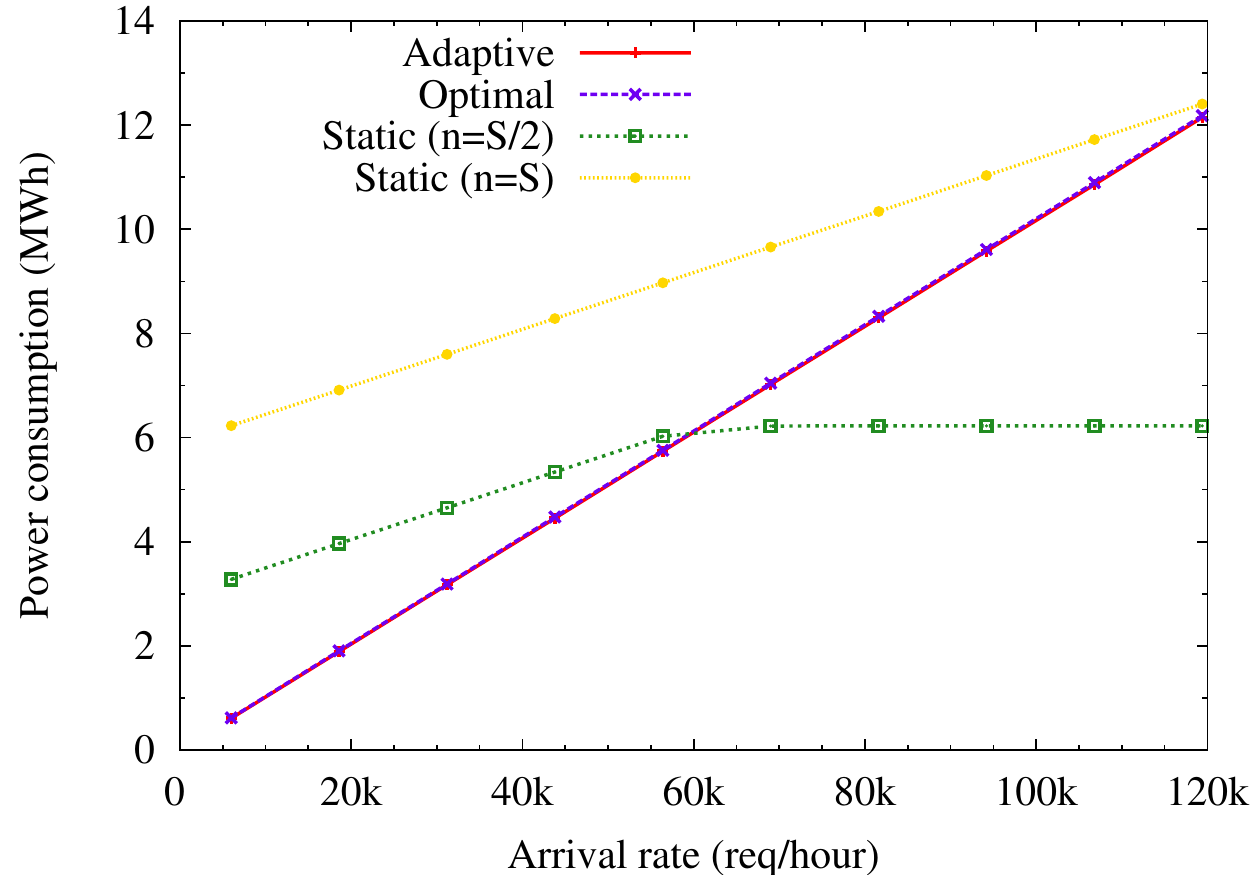}
\caption{Average power consumption for different policies.}
\label{fig:consumption}
\end{figure}

Finally, Figure~\ref{fig:consumption}, which depicts the
average power consumption, clearly shows that the dynamic policies
run servers only when needed, thus reducing the electricity bill and improving
the provider's profits.

Next, we depart from the assumption that the traffic is Markovian in order to
evaluate the effect of interarrival and service time variability on performance.
The average values are kept the same as before, however both the interarrival
and service times are generated according to a Log-Normal distribution. The
corresponding squared coefficient of variation are $ca^{2}=2$ and $cs^{2}=20$.
The high variability in job size distribution was deliberately chosen to
reflect the different kind of cloud users.

\begin{figure}[ht!]
\centering
\includegraphics[width=0.48\textwidth]{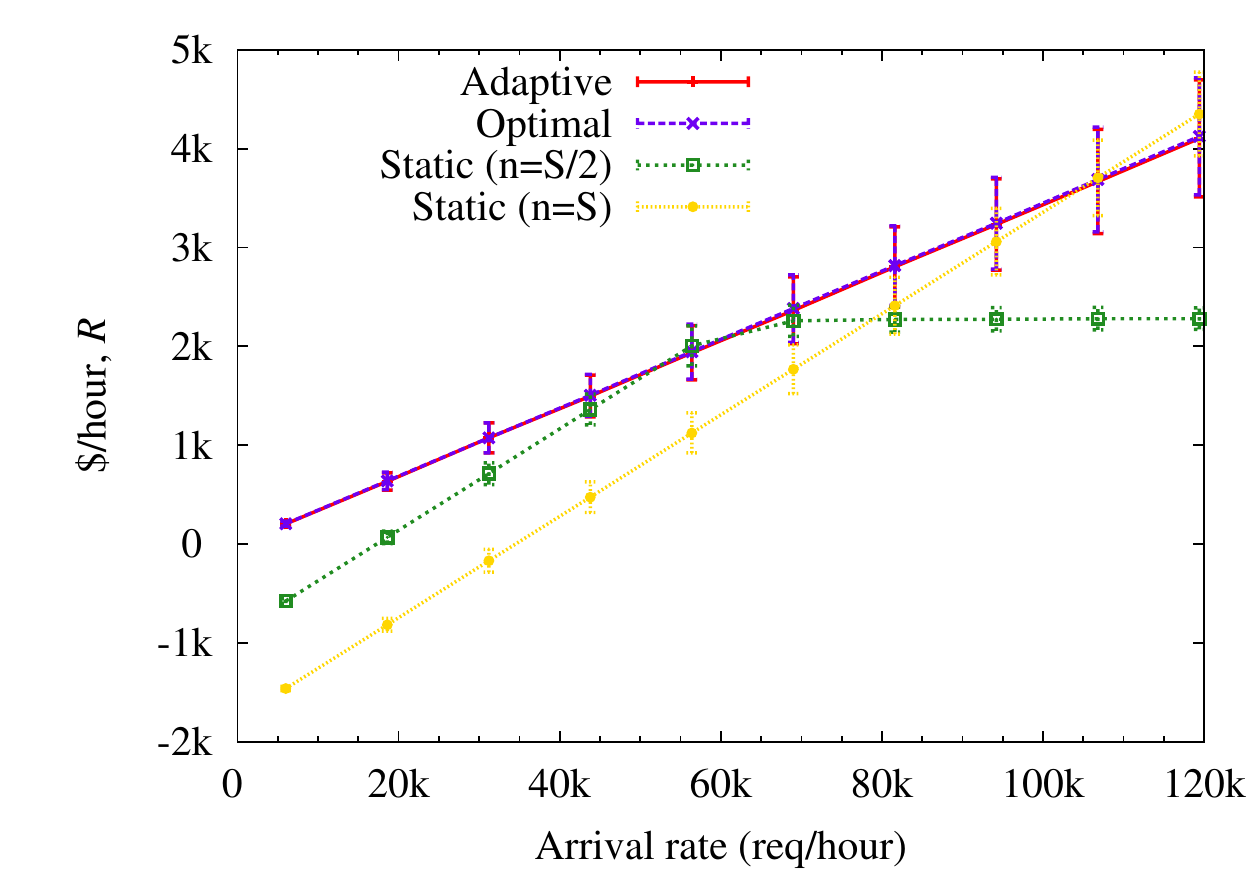}
\caption{Observed revenues for different policies, $ca^{2}=2$ and $cs^{2}=20$.}
\label{fig:ca_2_2_cs_2_20}
\end{figure}

It is legitimate to expect the performance to deteriorate when the
traffic variability increases, since the system becomes less predictable and it
is more difficult to choose the best $n$.
In fact, Figure~\ref{fig:ca_2_2_cs_2_20} shows that the achieved revenues are indeed
lower than those achieved when the traffic is Markovian.

%Sometimes, values other than the average revenue earned per unit time
%might be of interest. A possible example is the average power consumption, in
%case one wishes to limit the amount of energy drawn by the server
%farm. The average energy needed by the server farm under different policies
%plotted in Figure~\ref{fig:consumption} against the offered load. 
%The results show that the proposed policies run only the necessary amount  of
%servers (the power consumption grows roughly linearly with the load).

%\begin{figure}[ht!]
%\centering
%\includegraphics[width=0.45\textwidth]{figures/experiments/erlangB/consumption}
%\caption{Observed power consumption for different policies.}
%\label{fig:consumption}
%\end{figure}

%\begin{figure}[ht!]
%\centering
%\includegraphics[width=0.48\textwidth]{figures/experiments/erlangB/power_vs_loss}
%\caption{Power consumption and job loss as function of the offered load.}
%\label{fig:erlangB_power_vs_loss}
%\end{figure} 

In the next series of the experiments we evaluate the performance of the
proposed policies under non stationary loading conditions. Unfortunately, 
there is no publicly available data describing the demand for Cloud resources,
and thus we extrapolated it from the available Wikipedia
traces~\cite{urdaneta:2009}. 
%We speculate that Wikipedia traffic shape is 
%typical for a lot of applications hosted in the Cloud. 
Therefore the increase/decrease in the Wikipedia traffic would correlate  with
the general increase/decrease of the request rate for the resources. 
The arrival rate behavior has a general trend, with monthly, weekly and daily 
patterns, as well as unexpected spikes, which are hard to predict. 
We believe that such a workload is unbiased and thus will not provide
advantages for any specific approach. 
We assume that jobs enter the system arriving according to a Poisson process
with a certain rate $\lambda$ which changes every hour, while the system is
reconfigured every 30 minutes. 
%The datacenter contains 25,000 of four core boxes which total in 100,000 cores.

\begin{figure}[ht!]
\centering
\includegraphics[width=0.48\textwidth]{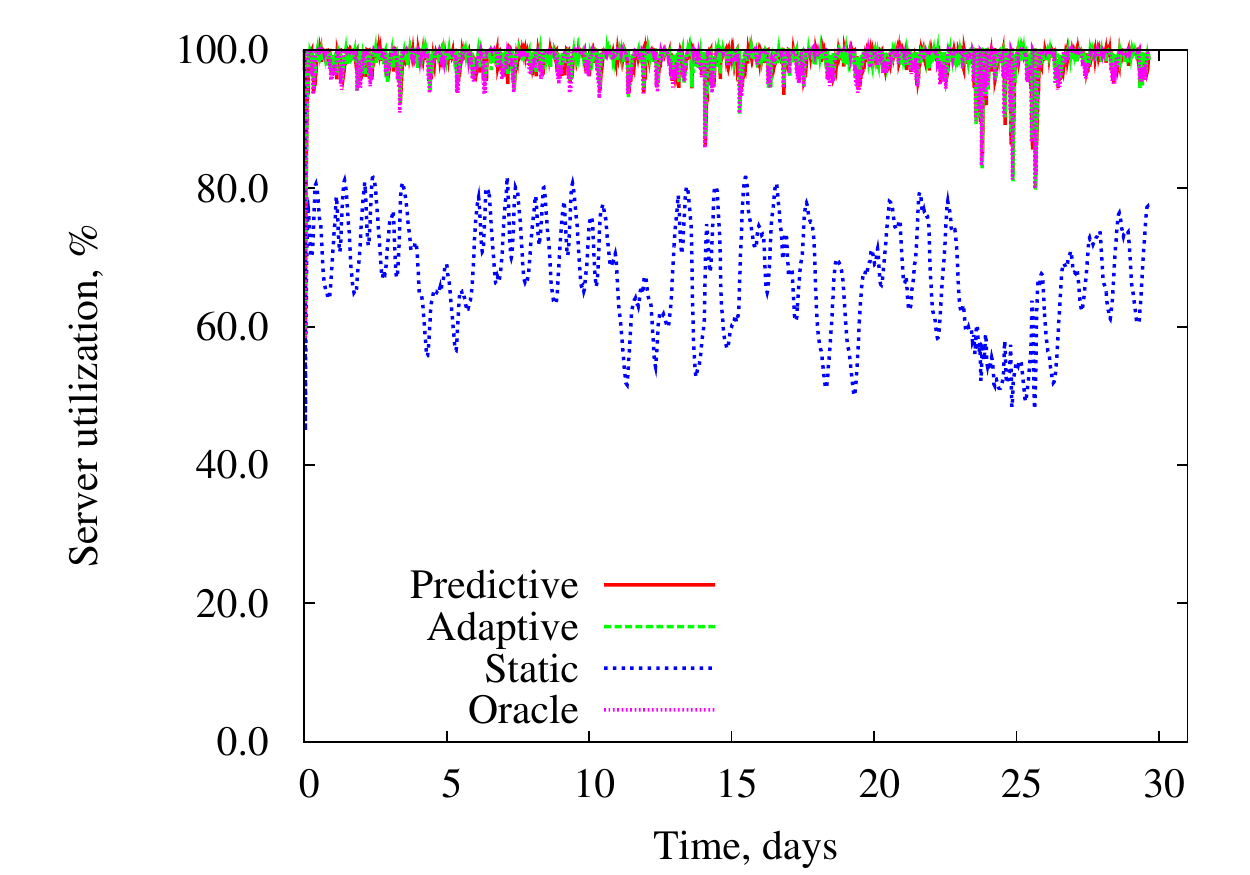}
\caption{Data center utilization.}
\label{fig:server_utilization}
\end{figure}

As it has been pointed out above, the QED algorithm performs almost as good as
the optimal allocation policy. Therefore, the next set of experiments are conducted
using the QED algorithm.
We evaluate the performance of QED with adaptive and predictive heuristic,
and, for comparison reasons, we also include a static allocation policy that
runs all the available servers, and an `Oracle' policy that knows the
exact value of $\lambda$ for the next time interval, and thus 
allocates the optimal number of servers.
Figure \ref{fig:server_utilization} shows that the static allocation policy
achieves 55\%-80\% utilization, while the dynamic policies use a smaller amount
of servers for handling the load, thus the achieved utilization is significantly higher.

\begin{figure}[ht!]
\centering
\includegraphics[width=0.48\textwidth]{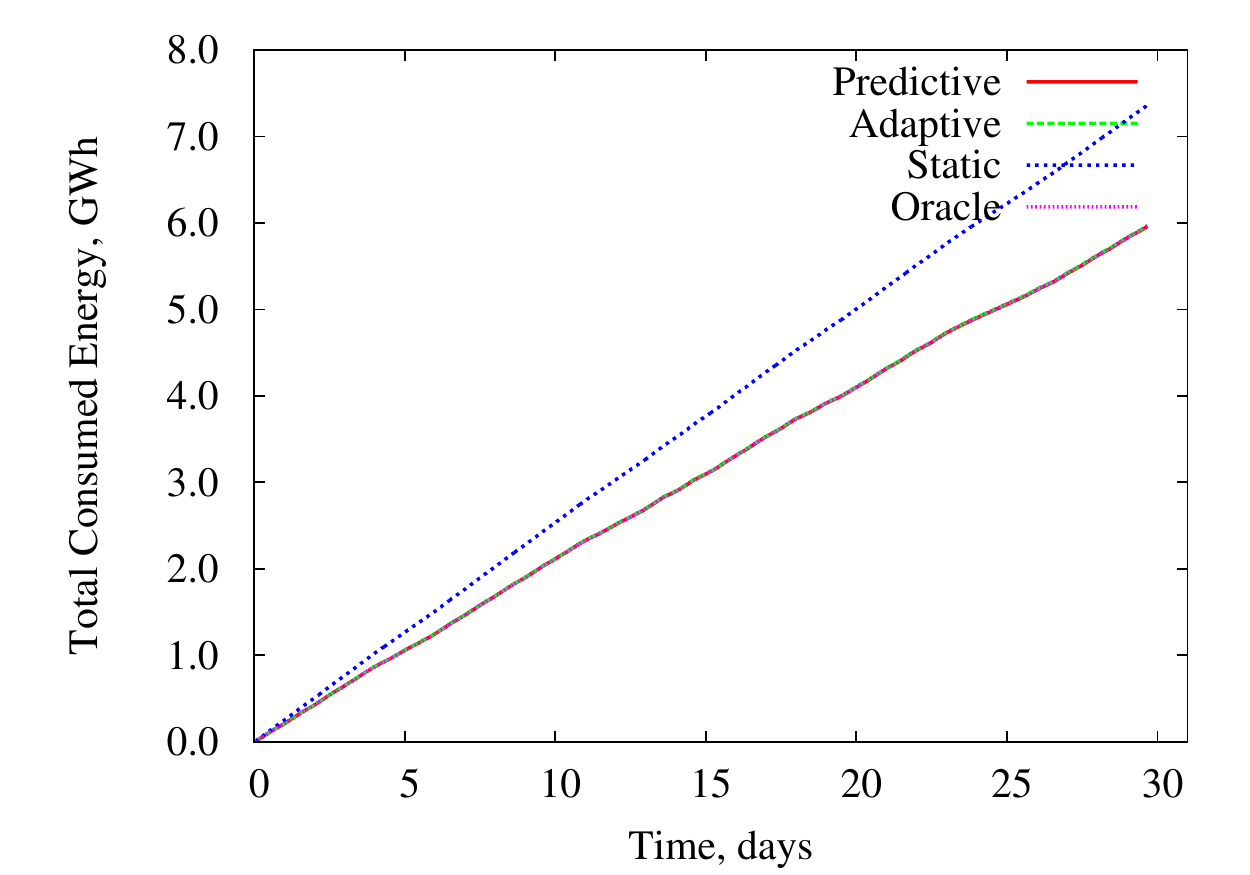}
\caption{Cumulative power consumption measured over one month}
\label{fig:cumulative_power}
\end{figure} 

\begin{figure}[ht!]
\centering
\includegraphics[width=0.48\textwidth]{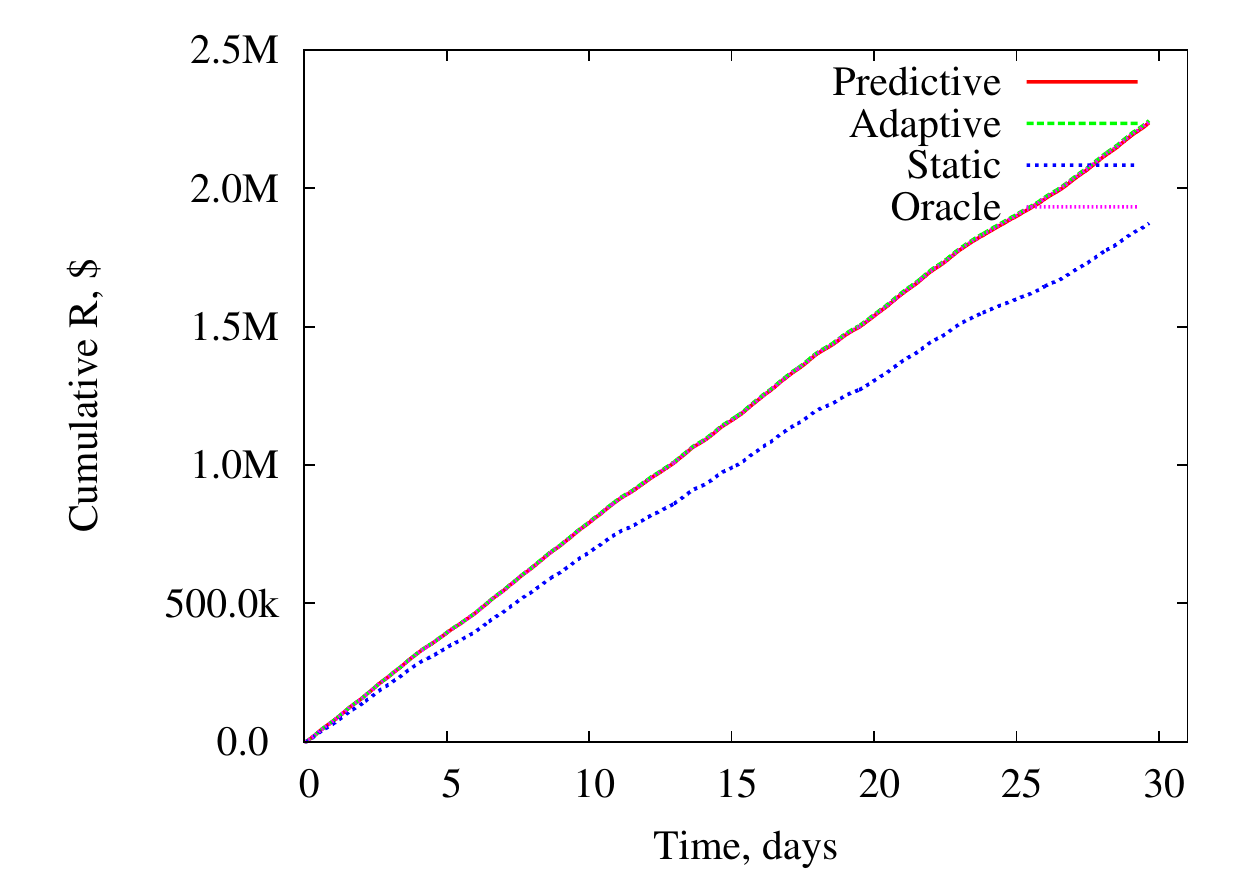}
\caption{Cumulative revenue measured over one month.}
\label{fig:cumulative_revenue}
\end{figure}

Since fewer servers are needed for handling the same load, the resulting power
consumption is also markedly smaller [fig. \ref{fig:cumulative_power}].
It is interesting to observe that despite the difference in the prediction 
mechanisms, all QED variations demonstrate almost identical results in terms
of achieved cumulative revenue, see Figure~\ref{fig:cumulative_revenue}. This
can be attributed to the fact the load fluctuation within the reconfiguration
intervals  is rather small and hard to predict.

\section{Conclusions}
\label{sec:conclusions}

We have introduced and throughly evaluated easily implementable
policies for dynamically adaptable cloud provision.
We have demonstrated that
decisions, such as how many servers are powered on, can have a significant effect on the 
revenue earned by the provider.
Moreover, those decisions are affected by the 
contractual obligations between clients and provider. The experiments we have 
carried out showed that the proposed polices work well under different traffic 
conditions, and that the `Adaptive' heuristic would be a good candidate for 
practical implementation.

Possible directions for future research include taking into account the time
and energy consumed during systems reconfigurations, trade offs between the
number of running servers and the frequency of the CPUs, and the power consumed by the networking equipment ({\it i.e.}, switches).

\section*{Acknowledgements}

This work was partly funded by the European 
Commission under the Seventh Framework Programme through the SEARCHiN project 
(Marie Curie Action, contract no. FP6-042467). 
The authors would also like to thank EU Cost Action IC 0804.

%\begin{spacing}{0.7}
\bibliographystyle{IEEEtran}%\setlength{\itemsep}{-2mm}
%\bibliographystyle{latex8} \setlength{\itemsep}{-2mm}
%{\tiny
\bibliography{green-computing}
%}
%\end{spacing}

\end{document}